\documentclass[conference]{IEEEtran}
\usepackage{dcase2026}

\usepackage{bm} 

\usepackage{dcase2026,amsmath,graphicx,url,times,booktabs, tabularx}

\newcommand{\method}{BADGE-Greedy-DPP}
\newcommand{\naulc}{N\mbox{-}AULC}
\newcommand{\rnaulc}{Rare\mbox{-}N\mbox{-}AULC}
\newcommand{\initbudget}{300}
\newcommand{\roundbudget}{300}
\newcommand{\totalbudget}{3000}
\newcommand{\ntrials}{10}
\newcommand{\nclasses}{10}
\newcommand{\nframes}{20}
\newcommand{\framelen}{0.5}

\title{Greedy Volume Maximization of Gradient Embeddings for Long-Tailed Frame-Level Bioacoustic Active Learning}

\name{Shiqi Zhang\(^{1,2}\),
      Marius Faiß\(^{2,3,4}\),
      Ariana Strandburg-Peshkin\(^{2}\),
      Tuomas Virtanen\(^{1}\)}
\address{\(^{1}\)Audio Research Group, Tampere University, Tampere, Finland\\
\texttt{\{shiqi.zhang, tuomas.virtanen\}@tuni.fi}\\
\(^{2}\)Department for the Ecology of Animal Societies, Max Planck Institute of Animal Behavior\\
\(^{3}\)International Max Planck Research School for Quantitative Behavior, Ecology and Evolution\\
\(^{4}\)Department of Biology, University of Konstanz\\
Konstanz, Germany\\
\texttt{\{mfaiss, astrandburg\}@ab.mpg.de}
}

\begin{document}
\maketitle

\begin{abstract}
    Bioacoustic call-type classification relies on costly expert annotation. Active learning can reduce this burden by selecting a small batch of segments for expert annotation and using the labeled segments for training the classifier. The setting is hard: the target calls are extremely sparse and the call-type distribution is long-tailed, so a tight budget must be spent on the few rare, informative segments. We propose \method, a deterministic batch selector that greedily adds the segment whose BADGE gradient embedding most enlarges the volume spanned by the batch; because this log-volume objective is submodular, the greedy rule guarantees a batch value at least a \((1-1/e)\) fraction of the optimum of this objective, a guarantee not provided by BADGE's existing k-means++ and MCMC DPP sampling heuristics. There is also a temporal granularity mismatch in the task. The acquisition function scores whole segments, yet the informative frames inside them are few. Uniform averaging therefore washes them out. The BADGE construction naturally addresses this mismatch when applied frame-wise, as prediction residuals weight the aggregated pseudo-gradient, so confidently predicted no-call frames contribute little while a single uncertain rare-call frame can still set the segment's direction. Across \ntrials\ runs on a sparse, imbalanced hyena call-type dataset, \method\ achieves the best overall and rare-call-type performance among all compared query strategies, including MFFT, the strongest non-BADGE baseline, and the two vanilla BADGE traversals.
\end{abstract}

\begin{IEEEkeywords}
    active learning, determinantal point process, frame-level audio classification, bioacoustics, imbalanced classes
\end{IEEEkeywords}

\section{Introduction}
\label{sec:intro}

Manual annotation of bioacoustic datasets is expensive, especially when recordings need temporally strong annotations with onset and offset rather than only clip-level tags \cite{stowell2022bioacoustics}. Active learning can reduce this effort by selecting a small batch of unlabeled audio segments for expert annotation before retraining the classifier \cite{zhao2020sed,martinsson2024weakstrong,wang2022activefewshot,shishkin2021mcdropout,shishkin2024bayesian,lindsey2023online,mcewen2024activefewshot,lindholm2025aggregation}. The difficulty in bioacoustics is that the pool is both sparse and long-tailed. Most frames contain no target sound, and the call types are imbalanced. Studies under long-tailed labels further show that minority categories stay poorly covered at tight label budgets \cite{tomanek2009imbalance,choi2021vabal,cui2019classbalanced,bengar2022classbalanced,zhang2018oa3,zhu2024generative,zhang2023longtailed}, so a useful query strategy must keep finding the few segments that carry rare, informative content rather than re-sampling the abundant background.

Most acquisition functions score a candidate by prediction uncertainty, by disagreement among a committee of models, by coverage of the feature space, or by explicit batch diversity \cite{lewis1994sequential,settles2012active,seung1992qbc,freund1997qbc,gonzalez1985farthest,sener2018coreset,gal2017deepbayesian,beluch2018ensembles,kirsch2019batchbald,citovsky2021batchscale,biyik2019dppbatch}. Batch active learning by diverse gradient embeddings (BADGE) instead unifies uncertainty and diversity by representing each candidate with a single gradient embedding built from the model prediction \cite{ash2020badge}. These methods usually assume that each query candidate can be represented by one vector. Frame-level audio classification breaks this granularity assumption: the learner queries a whole segment, but each segment contains many frames, most of which are silent, non-target, or confidently predicted, while only a few remain uncertain. Directly averaging them into a segment vector lets the many uninformative frames wash out the few informative ones, diluting the frame-level feature that should drive acquisition.

Our first contribution is a deterministic traversal for BADGE gradient-space batch selection. Original BADGE constructs each batch with either k-means++ seeding or finite-scan Markov chain Monte Carlo sampling from a fixed-size determinantal point process (MCMC k-DPP) \cite{arthur2007kmeanspp,anari2016mcmc}. Both rules tend to select diverse samples with large gradient-embedding norms, but they are sampling heuristics and do not give a worst-case lower bound on the quality of the selected batch. We instead use greedy selection, the standard mode-finding routine for DPPs \cite{kulesza2012dpp,biyik2019dppbatch}, to maximize the regularized log-determinant of the selected segment embeddings, which is actually the log-volume they span in gradient space \cite{kulesza2012dpp,krause2008sensor}. Because this log-volume objective is monotone and submodular, at a fixed batch size the greedy traversal returns a batch whose value is at least a \((1-1/e)\) fraction of the optimum for this uncertainty-diversity proxy objective \cite{nemhauser1978submodular}.

Our second contribution adapts the BADGE pseudo-gradient representation to frame-level audio active learning. Instead of uniformly averaging frame features, we construct and aggregate frame-wise pseudo-gradients weighted by prediction residuals. This maps each segment into BADGE gradient space. Because residual magnitudes are large near the decision boundary and close to zero on confidently predicted no-call frames, a few uncertain frames, even a single uncertain rare-call frame, can determine the segment's gradient direction. The resulting representation aggregates frames by informativeness rather than by frame count, so it is better suited to sparse, long-tailed frame-level queries.

We evaluate on an extremely sparse and imbalanced spotted-hyena frame-level call-type classification dataset \cite{woerner2026hyenaset}. All methods use fixed animal2vec embeddings \cite{schaferzimmermann2024animal2vec} and a 2-layer MLP head. Across \ntrials\ independent runs, \method\ achieves the best overall and rare-call-type active-learning quality among all compared strategies, including the strongest non-BADGE baseline, mismatch-first farthest traversal (MFFT) \cite{zhao2018mfft,zhang2025mfftbioacoustic}, and two vanilla BADGE traversals. Its margin is largest on the rarest classes, consistent with the exploration hypothesis behind the greedy volume objective.

\section{Method}
\label{sec:method}

\subsection{Active-Learning Framework and BADGE Embeddings}

We work in a standard pool-based active-learning loop. Let \(L_r\) be the labeled set of audio segments and \(U_r\) the unlabeled pool at annotation round \(r\). Each segment is a 10\,s clip divided into \(T\) fixed-length frames, and each frame may carry any subset of \(\nclasses\) call-type labels (multi-label). A classifier trained on \(L_r\) predicts these frame labels. The query rule ranks the unlabeled segments in \(U_r\), and the top \(B\) segments (the per-round budget) are labeled by an expert at the frame level and moved from \(U_r\) to \(L_{r+1}\). A useful query rule prioritizes segments whose labels should improve the next classifier, either because the current prediction is uncertain or because the segment covers a poorly represented region.

For segment \(i\) and frame \(t\), the classifier produces a posterior vector \(\mathbf{p}_{it} \in [0,1]^{\nclasses}\) and a penultimate-layer feature vector \(\mathbf{h}_{it}\). Following BADGE \cite{ash2020badge}, we form a pseudo label (termed a hallucinated label in the original BADGE study) by thresholding the current prediction as
\begin{equation}
    \hat{\mathbf{y}}_{it} = \mathbb{I}\left[\mathbf{p}_{it} > 0.5\right],
\end{equation}
where \(\mathbb{I}[\cdot]\) is the element-wise indicator, equal to one where the posterior exceeds \(0.5\) and zero otherwise. This pseudo label only builds the query representation; it is not a training target. We use it as the temporary label in a frame-wise pseudo binary-cross-entropy loss and take the gradient with respect to the final classification-layer weights. For a sigmoid multi-label head, this frame-wise gradient has the form \((\mathbf{p}_{it}-\hat{\mathbf{y}}_{it}) \otimes \mathbf{h}_{it}\). We aggregate these frame-wise gradient embeddings as
\begin{equation}
    \mathbf{G}_i = \frac{1}{T} \sum_{t=1}^{T} \left(\mathbf{p}_{it} - \hat{\mathbf{y}}_{it}\right) \otimes \mathbf{h}_{it},
    \label{eq:badge}
\end{equation}
where \(\otimes\) is the outer product. We flatten \(\mathbf{G}_i\) into a vector \(\boldsymbol{\phi}_i = \mathrm{vec}(\mathbf{G}_i)\), the segment representation used for selection. Unlike a segment-wise gradient built from a pooled segment feature and a segment-presence pseudo label, which assigns one residual to the whole segment and is blind to where frame-level uncertainty occurs, \eqref{eq:badge} keeps a separate residual for each frame. The residual \(\mathbf{p}_{it}-\hat{\mathbf{y}}_{it}\) is largest near the decision threshold, and \(\mathbf{h}_{it}\) records where that uncertainty sits in feature space, so each frame enters the sum in \eqref{eq:badge} weighted by its own residual magnitude. The aggregation over frames is therefore uncertainty-weighted rather than a uniform mean. Boundary frames dominate \(\mathbf{G}_i\), while frames the model predicts confidently (residual near zero), including the abundant no-call ones, contribute almost nothing. A segment with only a few informative frames thus keeps a distinctive embedding direction. The factor \(1/T\) scales every frame alike and serves only to keep the log-determinant objective \eqref{eq:logdet} in a stable numerical range; it does not change the relative weighting among frames. Acquisition then operates on these per-segment embeddings to bound query-stage compute and candidate-embedding storage, while predictions and supervision stay frame-level. We change only the traversal rule applied to these embeddings.

\subsection{Greedy DPP Traversal in Gradient Space}

\begin{figure*}[!t]
    \centering
    \includegraphics[width=\textwidth]{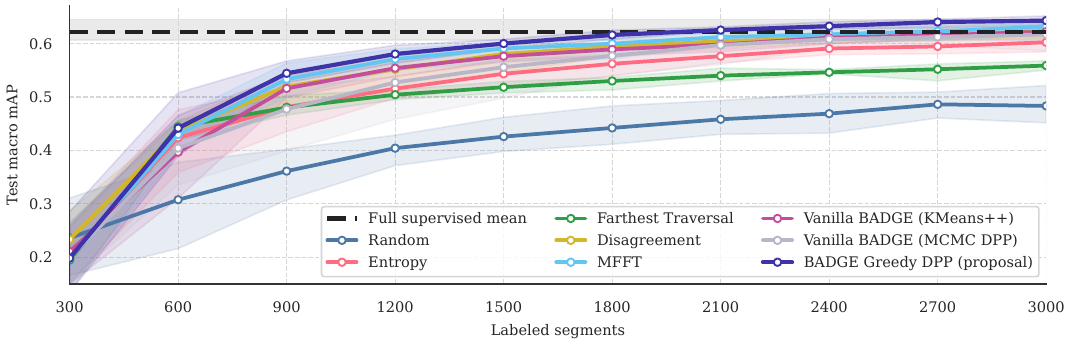}
    \caption{Active-learning curves on the hyena frame-level call-type task. Lines show mean test mAP across \ntrials\ runs. Colored bands show min-max ranges across runs, not confidence intervals. The dashed black line and gray band show the mean and min-max range of the separate full-supervised reference.}
    \label{fig:curves}
\end{figure*}

The traversal aims to select a batch of unlabeled segments whose gradient embeddings span a large volume. For a candidate selected set \(S\), let \(\boldsymbol{\Phi}_S\) be the matrix whose rows are the vectors \(\{\boldsymbol{\phi}_j : j \in S\}\). We score the selected set with the regularized log-determinant objective
\begin{equation}
    F(S) = \log \det \left( \mathbf{I} + \lambda^{-1} \boldsymbol{\Phi}_S^\top \boldsymbol{\Phi}_S \right),
    \label{eq:logdet}
\end{equation}
where \(\lambda > 0\) is a small regularizer. Up to a constant, \(F(S)\) is the log-volume spanned by the selected gradient embeddings, the quantity a DPP rewards \cite{kulesza2012dpp}. We build the batch greedily, at each step adding
\begin{equation}
    i^\star = \arg\max_{i \notin S} \Delta(i \mid S),
\end{equation}
with marginal gain
\begin{align}
    \Delta(i \mid S)
     & = F(S \cup \{i\}) - F(S) \nonumber                                                          \\
     & = \log\!\left(1 + \boldsymbol{\phi}_i^\top
               \left(\lambda \mathbf{I} + \boldsymbol{\Phi}_S^\top \boldsymbol{\Phi}_S\right)^{-1}
    \boldsymbol{\phi}_i\right),
    \label{eq:marginal}
\end{align}
obtained from the matrix determinant lemma.

Equation \eqref{eq:marginal} is the log-volume increase from adding candidate \(i\) to the current batch. It has a direct geometric reading: a candidate scores high when its gradient still has a large component outside the span of the already selected gradients, measured by the quadratic form \(\boldsymbol{\phi}_i^\top\left(\lambda \mathbf{I} + \boldsymbol{\Phi}_S^\top \boldsymbol{\Phi}_S\right)^{-1}\boldsymbol{\phi}_i\). The selector therefore favors segments that expand the span of the selected gradients over those that reinforce directions already well covered. We call this behavior \emph{exploration in BADGE gradient space}. In addition, \(F\) is monotone and submodular in the selected set, with \(F(\varnothing)=0\) \cite{krause2008sensor,nemhauser1978submodular,wei2015submodularity}; at a fixed batch size, the deterministic greedy batch therefore attains at least a \((1-1/e)\) fraction of the optimum of this regularized log-volume proxy for batch uncertainty and diversity, a worst-case bound. Moreover, because BADGE already encodes uncertainty through the pseudo-gradient construction~\cite{ash2020badge}, no additional weight between uncertainty and exploration is needed, unlike in DPP batch selection with a separate uncertainty score~\cite{biyik2019dppbatch}. At each round, we train the classifier with the same architecture on the current labeled set, compute \eqref{eq:badge} for all unlabeled segments, select \(B\) segments greedily with \eqref{eq:marginal}, and request their labels.

\section{Experimental Setup}
\label{sec:setup}

\begin{table}[!t]
    \centering
    \caption{Train-pool prevalence of \nclasses\ call types, sorted by segment prevalence. Segment prevalence is the fraction of 10\,s segments with at least one frame of the type; frame prevalence is the fraction of \framelen\,s frames carrying it.}
    \label{tab:prevalence}
    \renewcommand{\arraystretch}{1.05}
    \resizebox{0.85\linewidth}{!}{%
        \begin{tabular}{lcc}
            \toprule
            Call type           & Segment (\%) & Frame (\%) \\
            \midrule
            feeding (fed)       & 8.267        & 5.887      \\
            regular groan (grn) & 4.194        & 1.204      \\
            other (oth)         & 3.978        & 0.676      \\
            whoop (whp)         & 2.398        & 0.935      \\
            squeal (sql)        & 1.847        & 0.413      \\
            giggle (gig)        & 1.408        & 0.259      \\
            alarm rumble (rum)  & 1.273        & 0.303      \\
            squitter (str)      & 0.771        & 0.219      \\
            snore (snr)         & 0.444        & 0.247      \\
            growl (gwl)         & 0.425        & 0.072      \\
            \bottomrule
        \end{tabular}}
\end{table}

\begin{table*}[!t]
    \centering
    \caption{Main active-learning results over \ntrials\ runs; metrics are defined in Section~\ref{subsec:metrics}. \naulc, \rnaulc, F-mAP, and F-rmAP are percentages, and QT is in seconds. `*' averages reaching runs only, and `--' means no run reached the reference. Bold marks the best effectiveness values, highest FS-Rch, and lowest FS-Bud.}
    \label{tab:main}
    \renewcommand{\arraystretch}{1.07}
    \resizebox{\textwidth}{!}{
        \begin{tabular}{lcccccccc}
            \toprule
            Method                               & \naulc                    & \rnaulc                   & F-mAP                     & F-rmAP                    & R-Enr                       & FS-Bud                    & FS-Rch         & QT                   \\
            \midrule
            Random                               & 41.3 \(\pm\) 1.8          & 24.5 \(\pm\) 2.9          & 48.3 \(\pm\) 2.2          & 33.5 \(\pm\) 3.2          & 1.03 \(\pm\) 0.09           & --                        & 0\%            & 0.00 \(\pm\) 0.00    \\
            Entropy                              & 52.1 \(\pm\) 1.2          & 38.0 \(\pm\) 3.2          & 60.3 \(\pm\) 1.1          & 50.7 \(\pm\) 2.5          & 4.57 \(\pm\) 0.21           & --                        & 0\%            & 115.60 \(\pm\) 19.65 \\
            Farthest Traversal                   & 49.9 \(\pm\) 0.5          & 38.5 \(\pm\) 0.6          & 55.9 \(\pm\) 0.7          & 44.3 \(\pm\) 1.6          & 2.32 \(\pm\) 0.09           & --                        & 0\%            & 125.16 \(\pm\) 3.60  \\
            Disagreement                         & 54.9 \(\pm\) 0.7          & 40.9 \(\pm\) 2.3          & 62.4 \(\pm\) 0.8          & 53.7 \(\pm\) 2.0          & 8.59 \(\pm\) 0.17           & 2880 \(\pm\) 164*         & 50\%           & 355.32 \(\pm\) 19.51 \\
            MFFT                                 & 55.5 \(\pm\) 0.7          & 43.6 \(\pm\) 2.2          & 63.2 \(\pm\) 0.6          & 55.4 \(\pm\) 2.0          & 8.72 \(\pm\) 0.24           & 2533 \(\pm\) 339*         & 90\%           & 427.46 \(\pm\) 78.40 \\
            Vanilla BADGE (KMeans++)             & 54.2 \(\pm\) 1.0          & 43.7 \(\pm\) 1.8          & 62.3 \(\pm\) 0.4          & 55.5 \(\pm\) 1.1          & 5.74 \(\pm\) 0.33           & 2880 \(\pm\) 164*         & 50\%           & 125.30 \(\pm\) 40.12 \\
            Vanilla BADGE (MCMC DPP)             & 53.0 \(\pm\) 1.8          & 40.9 \(\pm\) 3.2          & 61.7 \(\pm\) 1.2          & 53.8 \(\pm\) 2.8          & 5.36 \(\pm\) 0.64           & 2925 \(\pm\) 150*         & 40\%           & 768.99 \(\pm\) 17.86 \\
            \textbf{BADGE Greedy DPP (proposal)} & \textbf{56.7 \(\pm\) 0.9} & \textbf{47.4 \(\pm\) 1.7} & \textbf{64.3 \(\pm\) 0.6} & \textbf{58.6 \(\pm\) 1.3} & \textbf{10.41 \(\pm\) 0.26} & \textbf{2040 \(\pm\) 126} & \textbf{100\%} & 141.96 \(\pm\) 9.77  \\
            \bottomrule
        \end{tabular}}
\end{table*}

We evaluate on a hyena frame-wise multi-label call-type classification task built from a preliminary version of the HyenaSET dataset \cite{woerner2026hyenaset}. The data are tracking-collar recordings from a spotted hyena clan of 19 spotted hyenas (\textit{Crocuta crocuta}) in the Masai Mara National Reserve in Kenya, collected by the Mara Hyena Project, Michigan State University, and the Max Planck Institute of Animal Behavior, and contain significant collar and environmental noise. We use the fully annotated sections, where call type, onset, and offset are marked, giving about 205 hours of audio with \nclasses\ call types. The recordings are cut into non-overlapping 10\,s segments, each split into \nframes\ frames of \framelen\,s.

The task is sparse and long-tailed. Under \(10\%\) of frames contain an annotated call. Table~\ref{tab:prevalence} reports the segment and frame prevalence of every call type; both span a strong long tail, from the most common type down to the three rarest, growl (\texttt{gwl}), snore (\texttt{snr}), and squitter (\texttt{str}).

We split the segments into train, validation, and test sets in a 70/15/15 ratio, stratified by segment-level call-type label. Each method is evaluated over \ntrials\ independent runs with different random seeds. A run starts from a randomly sampled seed set of \initbudget\ labeled segments and adds nine active-learning rounds of \(B = \roundbudget\) segments each, reaching a final labeled budget of \totalbudget\ segments.

\subsection{Representation, Model, and Compared Methods}

Following common active-learning practice, we freeze the audio representation and vary only the query strategy. All methods use fixed animal2vec embeddings \cite{schaferzimmermann2024animal2vec} and a 2-layer MLP head. This isolates the selection rule from representation-learning effects. The classifier is trained with binary cross-entropy loss and Adam, with model selection and early stopping on validation macro average precision (mAP).

The compared query strategies are Random, Entropy, Farthest Traversal, Disagreement, MFFT \cite{zhao2018mfft,zhang2025mfftbioacoustic}, vanilla BADGE (k-means++) \cite{ash2020badge,arthur2007kmeanspp}, vanilla BADGE (MCMC DPP) \cite{ash2020badge,anari2016mcmc}, and \method. Entropy, Farthest Traversal, and Disagreement operate on the same frame-level predictions, and MFFT applies mismatch-first farthest traversal. All BADGE variants use the same pseudo-gradient embedding construction in \eqref{eq:badge}; they differ only in the traversal step, which selects a batch by k-means++, MCMC DPP, or our greedy log-det rule. For \method, we fix the regularizer to \(\lambda = 10^{-6}\) as a small ridge term for invertibility and numerical stability. We also train a full-supervised reference model that uses the same protocol (data splits, animal2vec representation, MLP head, loss, optimizer, early-stopping rule, and test metric).

\begin{figure*}[!t]
    \centering
    \includegraphics[width=\textwidth]{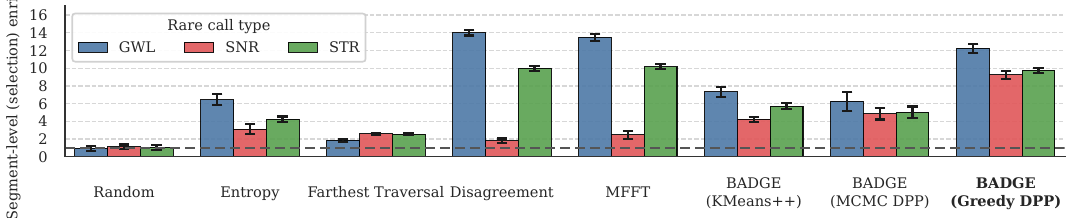}
    \caption{Final segment-level enrichment of the three rare call types relative to their train-pool prevalences, compared across all query strategies. The dashed line at 1.0 corresponds to prevalence-matched selection. Error bars show standard deviation over \ntrials\ runs.}
    \label{fig:enrichment}
\end{figure*}

\subsection{Metrics}
\label{subsec:metrics}

Our primary metric is \naulc, which summarizes the whole learning curve instead of only the last labeled budget. We compute \naulc\ as
\begin{equation}
    \naulc =
    \frac{1}{b_9 - b_0}
    \sum_{k=1}^{9}
    (b_k - b_{k-1})\frac{m_k + m_{k-1}}{2},
\end{equation}
where \(b_k = \initbudget + k \cdot \roundbudget\) is the cumulative labeled budget after round \(k\) for \(k=0,\ldots,9\), and \(m_k\) is the test mAP at round \(k\). Thus, \naulc\ is the budget-averaged test mAP over the learning curve and the higher the better. We compute \rnaulc\ analogously by applying the same formula to the rare-call-type mAP over \{\texttt{gwl}, \texttt{snr}, \texttt{str}\}.

At \totalbudget\ labeled segments, final mAP (F-mAP) and final rare-call-type mAP (F-rmAP) summarize endpoint performance. Rare macro enrichment (R-Enr) averages, over \texttt{gwl}, \texttt{snr}, and \texttt{str}, the ratio between selected-set and train-pool segment prevalence. FS-Bud records the first cumulative labeled budget when a run reaches the mean test mAP of the same-protocol full-supervised reference, and FS-Rch records the fraction of runs that do so. FS-Bud means and standard deviations use reaching runs only. QT sums the total query-stage wall-clock time over the nine active-learning rounds.

\section{Results}
\label{sec:results}

Table~\ref{tab:main} and Fig.~\ref{fig:curves} show that \method\ is the best active learner among the methods compared here. It attains the highest \naulc\ (\(56.7\% \pm 0.9\%\)), final mAP (\(64.3\% \pm 0.6\%\)), and rare-call-type metrics. The closest competitor is MFFT, which it exceeds by \(+1.1\) points in final mAP.

For reference, same-protocol full-supervised runs reach \(62.2\% \pm 1.5\%\) final mAP. Under one-sided, unpaired exact mean-difference permutation tests, Holm-corrected over the seven baselines per metric \cite{holm1979sequential}, \method\ improves over MFFT, vanilla BADGE (k-means++), and vanilla BADGE (MCMC DPP) by \(+1.2\), \(+2.5\), and \(+3.7\) points in \naulc\ (corrected \(p=1.60\times10^{-3}\), \(4.87\times10^{-5}\), and \(3.79\times10^{-5}\)), and by \(+3.8\), \(+3.7\), and \(+6.5\) points in \rnaulc\ (corrected \(p=3.57\times10^{-4}\), \(3.57\times10^{-4}\), and \(3.79\times10^{-5}\)).

\subsection{Rare Call-Type Performance}

The rare-class gains reflect more than stronger overall performance. In Table~\ref{tab:main}, \method\ has the highest \rnaulc, final rare-call-type mAP, rare macro enrichment, and final count of selected segments carrying at least one rare-call-type label (\(496.8 \pm 11.3\)).

Figure~\ref{fig:enrichment} shows the per-class pattern. Disagreement and MFFT over-sample \texttt{gwl} and \texttt{str} (mean enrichments of \(14.00\times\) and \(9.97\times\) for Disagreement, \(13.51\times\) and \(10.19\times\) for MFFT) but stay weak on \texttt{snr}, with mean \texttt{snr} enrichments of only \(1.81\times\) and \(2.46\times\). \method\ is more balanced across the tail, with \(12.24\times\), \(9.23\times\), and \(9.76\times\) enrichment for \texttt{gwl}, \texttt{snr}, and \texttt{str}; this is consistent with rewarding underrepresented directions in BADGE gradient space.

\subsection{Full-Supervised Reference and Efficiency Trade-Off}

\method\ is the only method that reaches the mean reference in all \ntrials\ runs (FS-Rch), doing so after \(2040 \pm 126\) labeled segments (FS-Bud; \(3.94\% \pm 0.24\%\) of the full train set); MFFT reaches it in 90\% of runs at \(2533 \pm 339\) segments (\(4.89\% \pm 0.66\%\)). \method\ exceeds the full-supervised reference in the tail, matching it on the seven common call types (\(66.8\%\) versus \(66.9\%\) final mAP) and surpassing it on the three rare ones (\(58.6\%\) versus \(51.0\%\)). It is only moderately slower than BADGE (KMeans++) (\(141.96 \pm 9.77\)~s versus \(125.30 \pm 40.12\)~s) and much faster than BADGE (MCMC DPP) (\(768.99 \pm 17.86\)~s).

\section{Conclusion}
\label{sec:conclusion}

We introduced \method\ for active learning on long-tailed, frame-level bioacoustic classification. It brings BADGE's gradient embedding to the frame level, where residual-weighted aggregation lets a few uncertain frames set each segment's selection direction. It then builds the query batch by deterministic greedy maximization of the spanned volume, a monotone submodular rule with an approximation guarantee. On the hyena task it delivers the best overall and rare-call-type active-learning quality, with the largest margins on the rarest classes. We leave further datasets to future work.

\clearpage
\IEEEtriggeratref{22}
\bibliographystyle{IEEEtran}
\bibliography{refs}

\end{document}